\documentclass[12pt,a4paper]{article}

\usepackage[english]{babel}
\usepackage{amsmath,amssymb,amsfonts,amsthm}
\usepackage[mathscr]{eucal}
\usepackage[all]{xy}
\usepackage{hyperref}
\usepackage{setspace}
\usepackage{upgreek}
\usepackage{hyperref}
\usepackage{cite}

\usepackage{times}
\usepackage{color}

\usepackage{indentfirst}

\allowdisplaybreaks

\begin{document}

\title{Chiral effects in classical spinning gas}
\author {M.A. Bubenchikov, D.S.Kaparulin, and O.D. Nosyrev}
\date{\small\textit{
Faculty of mathematics and mechanics, Tomsk State University,\\ Lenina ave. 36, Tomsk 634050, Russia}}

\maketitle

\begin{abstract}
\noindent We consider a statistical mechanics of rotating ideal gas consisting of classical non-relativistic spinning particles. The microscopic structure elements of the system are massive point particles with a nonzero proper angular momentum. The norm of proper angular momentum is determined by spin. The direction of proper angular momentum changes continuously. Applying the Gibbs canonical formalism for the rotating system, we construct the one-particle distribution function, generalising the usual Maxwell-Boltzmann distribution, and the partition function of the system. The model demonstrates a set of chiral effects caused by interaction of spin and macroscopic rotation, including the change of entropy, heat capacity, chemical potential and angular momentum. 
\end{abstract}

\section{Introduction}

The statistical mechanics of rotating systems is studied for many
years. Maxwell was the first who considered the issue of thermodynamic
equilibrium of rotating bodies \cite{JCM-1878}. Gibbs developed the
general formalism for description of macroscopic rotation in the framework of canonical ensembles
\cite{JWG-1902}. The modern exposition of the concept can be found
in the textbook \cite{LL}. The early works have one common feature: the particles
are considered as point objects in space, so
the motion in the field of centrifugal forces is an issue. In 1915,
Barnett experimentally discovered magnetization of uncharged
rotating body \cite{Barnett}. The inverse effect was described by
Einstein and de Haas \cite{EdH}. Observations of Barnett, Einstein and de Haas represent the first examples of chiral effects, the class of
phenomena, being associated with the interaction of spin angular
momenta of individual particles and macroscopic rotation. Later, it
has been recognized that there is complex of chiral phenomena, that
influence almost all the macroscopic characteristics of the body,
including heat capacity and chemical potential. For a review of the problem, we cite the recent article \cite{Fukushima} and references therein.

The current theory of chiral effects is developing in the framework
of formalism of quantum mechanics and quantum field theory at finite temperature
\cite{Vilenkin}, \cite{Fukushima1}, \cite{Fukushima}, \cite{Chernodub}, \cite{Chernodub1}, \cite{Xu}. 
This approach, being the most
fundamental, has several limitations. First, the spin takes half-integer
values in quantum theory, while the particle equations of motion have slightly different form for different spins. This prevents
the construction of statistical theory involving the
particles of all spins on the equal basis. The majority of studies
focus on the particles with spin 1/2 and 1, i.e. spinor and
vector field. The references are given above. The most recent articles consider 
hydrodynamics of rotating fluid with angular momentum \cite{Becc}, \cite{She}, \cite{Florkowski}, \cite{Bhadury}, but this is a macroscopic theory. The studies on statistical mechanics of systems of rotating particles with higher spins (greater than two) are almost absent in the literature. Second, the existence of a
well-defined quantum description of higher spin fields in the
presence of external fields can be an issue. The problem is most
difficult for the massive particles where the restrictions for
gravitational or electromagnetic couplings are known \cite{Kaparulin}, \cite{Cortese}, \cite{Boulanger}. This
prevents construction of theory chiral effects of higher spin
particles, when they are moving in general gravitational or
electromagnetic field. Finally, the atoms or molecules can have
nonzero orbital angular momentum in ground state \cite{LL}. Once these
systems are considered as the point particles with proper angular momentum, they have 
possible obstructions for description within the formalism of
quantum statistical mechanics by the mentioned above reasons.

The classical theory of spin is developing since the work of Frenkel
\cite{Frenkel}. The co-orbit method tells us that the classical spinning particle is a dynamical system on the
co-orbit of the Poincare group \cite{Kirillov}, \cite{Kostant}, \cite{Soureau}. 
In most of the current applications, the
dynamical system is considered as a point particle in space with a proper
angular momentum. The proper angular momentum serves as the classical analog of spin. The classical theory of spinning particles is free of the above-mentioned limitations. The model parameters are mass and spin. Both
the quantities change continuously. This makes it possible a
construction of universal model of spinning particle, which
considers all the combinations of mass and spin on the equal
basis \cite{Fryd}. The equations of motion are known for free
particles of arbitrary spin, in the electromagnetic and
gravitational fields of arbitrary configuration \cite{Universal}. No restrictions on the background appear. An
equivalent model with the vector parameterization of spin states is
considered in \cite{Deriglazov}. The use of classical spin concept
extends the class of systems that admits description of chiral
effects. The other benefit of the classical theory is its relative
simplicity. This makes it an ideal probing tool for computation of
first estimates of chiral effects.

In the current research, we develop the quasi-classical theory of
chiral effects in the non-relativistic classical spinning gas. The
microscopic structure elements of the system are point particles with a 
proper angular momentum. The direction of proper angular momentum is determined by the particle state, and its norm is
defined by spin. The phase space of the spinning particle is
$\mathbb{R}^6\times\mathbb{S}^2$. The phase space coordinates
include, besides the space coordinates and linear momenta, two
angular variables that determine the direction of spin. In contrast to the quantum case, the direction of classical spin
angular momentum changes continuously, so each point on
$\mathbb{S}^2$ represents a possible spin state. The gas is supposed
to be ideal, so there is no interaction between the particles. The
shape of the gas reservoir is chosen to be a cylinder, rotating along
the symmetry axis. The axis of rotation is supposed to be fixed. We
construct a one particle distribution function, generalizing the usual
Maxwell-Boltzmann distribution, and find the partition function. The
values of chiral effects are computed in classical and quantum case,
and compare the results.

The article is organized as follows. In the next section, we describe a model of classical spinning particle. It is given by a non-relativistic limit of theory \cite{Deriglazov}. The invariant
measure on the phase space is shown to be a product of invariant
measure on the space of translational degrees of freedom, and
invariant measure on the sphere. In Section 3, we construct a one-particle distribution function. We show that it decouples into the
product of three functions depending on the momenta, positions, and
angular variables. For spins, it is observed that
the particles momenta have a tendency of orientation along the vector of angular
velocity. In Section 4, we construct a thermodynamics of spinning degree of freedom. The
partition function and the values of chiral corrections to entropy, 
spin angular momentum, and chemical potential are found. In Section 5, we discuss the limitations of the model of classical spinning gas, and give the estimates for the magnitude of chiral effects in (quasi-)classical theory. The conclusion
summarizes results.

\section{Phase space}

We consider a massive classical non-relativistic point particle with
spin travelling in three-dimensional Euclidean space. The generalized
coordinates of the model include the particle position in space,
being determined by the radius-vector $x^i, i=1,2,3,$ and
another vector $\omega^i, i=1,2,3$ describing spin.
Such a choice of configuration space corresponds to the
vector model of the spinning particle \cite{Fryd}. The generalized momenta,
being canonically conjugated to the generalized coordinates $x^i$,
$\omega^i$ are denoted $p^i$,
$\pi^i$. The bold symbols $\boldsymbol{r}$, $\boldsymbol{\omega}$, $\boldsymbol{p}$, $\boldsymbol{\pi}$ stand for the three-vectors of generalized coordinates and generalized momenta.
The phase-space
variables are supposed to be functions of time $t$. 
The canonical
Poisson bracket on the phase space read
\begin{equation}\label{pbr}\begin{array}{c}\displaystyle
   \{x^i,x^j\}=\{x^i,\omega^j\}=\{x^i,\pi^j\}=\{\omega^i,p^j\}=\{\omega^i,\omega^j\}=\{p^i,\pi^j\}=0\,,\\[3mm]
   \{x^i,p^j\}=\{\omega^i,\pi^j\}=\delta^{ij}\,.
\end{array}\end{equation}
The variables of spin
sector are subjected to constraints,
\begin{equation}\label{constr}
    (\boldsymbol{\omega},\boldsymbol{\omega})=a^2,\qquad (\boldsymbol{\pi},\boldsymbol{\pi})=b^2, \qquad (\boldsymbol{\omega},\boldsymbol{\pi})=0\,.
\end{equation}
(The round brackets denote the scalar product of two $3d$ vectors with respect to
the Euclidean metric.) The positive real numbers $a$, $b$ are model
parameters. Relations (\ref{constr}) imply that the vectors
$\boldsymbol{\omega}$, $\boldsymbol{\pi}$ are normalized and
orthogonal to each other. The constraints ensure that the physical phase space of the theory is the fiber bundle 
$\mathbb{R}^6\times \mathbb{S}^2$. This is a space of states of a spinning particle.

The Hamiltonian action functional for the model reads
\begin{equation}\label{S-fucnt}
    S_0=\int \bigg\{(\boldsymbol{p},\dot{\boldsymbol{q}})+(\boldsymbol{\pi},\dot{\boldsymbol{\omega}})-H\bigg\}dt\,;
\end{equation}
\begin{equation}
    H=\frac{(\boldsymbol{p},\boldsymbol{p})}{2m}+\lambda_{\omega\omega}((\boldsymbol{\omega},\boldsymbol{\omega})-a^2)+\lambda_{\pi\pi}((\boldsymbol{\pi},\boldsymbol{\pi})-b^2)+\lambda_{\omega\pi}(\boldsymbol{\omega},\boldsymbol{\pi})\,.
\end{equation}
The dynamical variables are the phase-space coordinates $\boldsymbol{r}$, $\boldsymbol{\omega}$, $\boldsymbol{p}$, $\boldsymbol{\pi}$ and
Lagrange multipliers $\lambda_{\omega\omega}$, $\lambda_{\pi\pi}$,
$\lambda_{\omega\pi}$. The energy and total angular momentum of a single spinning particle
(modulo constraints (\ref{constr})) read
\begin{equation}\label{J-tot}
    \varepsilon=\frac{p^2}{2m}\,,\qquad\boldsymbol{j}=[\boldsymbol{r},\boldsymbol{p}]+[\boldsymbol{\omega},\boldsymbol{\pi}]\,.
\end{equation}
The round brackets denote the standard vector product in $3d$ Euclidean
space. The total angular momentum $\boldsymbol{j}$ is given by a sum of
the orbital angular momentum $[\boldsymbol{q},\boldsymbol{p}]$ and spin
angular momentum $[\boldsymbol{\omega},\boldsymbol{\pi}]$. In what
follows, we introduce a special notation for the spinning part of
angular momentum,
\begin{equation}\label{J-spin}
    \boldsymbol{j}_{sp}=[\boldsymbol{\omega},\boldsymbol{\pi}]\,.
\end{equation}
Throughout the article, we assume that the quantization of the model
corresponds to the irreducible representation of the Poincare group
with the mass $m$ spin $S$. In this case, the spin angular momentum
vector is normalized, with the norm
\begin{equation}\label{J-spin-norm}
    (\boldsymbol{j}_{sp},\boldsymbol{j}_{sp})\equiv a^2b^2=\hbar^2S(S+1)\,.
\end{equation}
($\hbar$ stands for the Planck constant). The formula fixes the product $ab$. For example, for an electron
with spin $1/2$, we have to assume that $ab=\sqrt{3}\hbar/2$. For a fixed value of $S$, the parameters $a$, $b$ still have a scaling ambiguity,
$a\mapsto a\lambda$, $b\mapsto b\lambda^{-1}$ with the nonzero real
$\lambda.$ The presence of accessory
parameter $\lambda$ distinguishing the models with one and the same value
of mass and spin has been noted in \cite{Universal}. The geometrical
meaning of these parameters in terms of space-time trajectory shape
is explained in \cite{WS}. The accessory parameters are not relevant
in the context of statistical mechanics because they do not alter the space of
spinning particle states. In what follows, we use a special notation 
\begin{equation}
    \Sigma=\sqrt{S(S+1)}\,.
\end{equation}
By construction, $\Sigma=ab/\hbar$. The quantity $\Sigma$ is determined by spin without ambiguity. The norm of the classical spin angular momentum (\ref{J-spin}) is $\hbar\Sigma$.

The crucial object of classical statistical mechanics is the
invariant measure on the phase space. The construction of the
measure is non-trivial due to presence of constraints
(\ref{constr}). Below, we give a heuristic solution to the problem. The dimensionless element of phase space
(having a sense of elementary number of quantum states) can be
represented as
\begin{equation}
    d\Gamma=d\Gamma_{tr}d\Gamma_{sp}\,.
\end{equation}
Here, $d\Gamma_{tr}$ is the elementary number of transnational
states, and $d\Gamma_{sp}$ plays the same role for spin states. The
expression for $d\Gamma_{tr}$ is well-known,
\begin{equation}
    d\Gamma_{tr}=\frac{d\boldsymbol{r}d\boldsymbol{p}}{(2\pi\hbar)^3}\,.
\end{equation}
The construction of the volume element in the space of spinning sates is
quite tricky. At first, the constraints (\ref{constr}) can be solved
using the Euler angles $\alpha$, $\beta$, $\gamma$. By definition,
we put
\begin{equation}\begin{array}{c}
        \omega^1=a(\cos (\alpha)\cos(\gamma)-\cos(\beta)\sin(\alpha)\sin(\gamma)),\\[3mm] \omega^2=a(\cos(\alpha)\cos(\beta)\sin(\gamma)+\cos(\gamma)\sin(\alpha)),\qquad
        \omega^3=a\sin(\beta)\sin(\gamma)\,;
\end{array}\end{equation}
\begin{equation}\begin{array}{c}
    \pi^1=-b(\cos(\alpha)\sin(\gamma)-\cos(\beta)\cos(\gamma)\sin(\alpha)),\\[3mm]
    \pi^2=b(\cos(\alpha)\cos(\beta)\cos(\gamma)-\sin(\alpha)\sin(\gamma)),\qquad\pi^3=b\cos(\gamma)\sin(\beta))\,.
\end{array}\end{equation}
The range of new variables is
\begin{equation}
    0\leq\alpha\leq 2\pi,\qquad 0\leq\beta\leq \pi,\qquad 0\leq\gamma \leq 2\pi\,.
\end{equation}
($\pi$ stands for $\pi$-number.) The related invariant measure on
the space of possible configurations of vectors is the Haar measure
on the rotation group. This is \textit{not} a measure on the
space of spin states because the particle state, being determined by
the momentum and total angular momentum, does not depend on the
angle $\gamma$,
\begin{equation}\label{spin-vect}\begin{array}{c}
    \boldsymbol{j}_{sp}=\hbar\Sigma(\sin(\alpha)\sin(\beta),-\cos(\alpha)\sin(\beta),\cos(\beta))\,.
\end{array}\end{equation}
This means that the phase-space points with one and the same value of
variable $\gamma$ describe one physical state. To exclude
integration of over the angle $\gamma$, the gauge fixing $\gamma=0$
can be applied. The latter means that the invariant measure is
proportional to the volume element sphere,
\begin{equation}
    d\Gamma_{sp}\sim\sin(\beta)d\alpha d\beta\,.
\end{equation}
An overall constant is a matter of convention because the volume element has dimension in classical theory. We chose the following
normalizing condition:
\begin{equation}
    \int d\Gamma_{sp}=2\Sigma\,.
\end{equation}
It means that $2\Sigma$ quasiclassical states are accessible for the
particles with the classical spin $\Sigma$. For the big values of spin
$S$, we have estimate $2\Sigma=2S+1+O(1/S)$. This mimics the well-known fact: the quantum particle with spin $S$ has $2S+1$ spin
states. The final result for invariant measure on the space of
spinning particle states reads,
\begin{equation}\label{dg}
    d\Gamma=\frac{\Sigma}{2\pi(2\pi\hbar)^3}\sin(\beta)d\boldsymbol{r}d\boldsymbol{p}d\alpha d\beta\,.
\end{equation}
In what follows, we use this formula.

The obtained invariant measure admits a simple explanation. In the
recent article \cite{WS}, it has been shown that the configuration
space of the spinning particle can be realized in the form of set of
cylindrical (hyper) surfaces in Minkowski space. The world sheet of
massive spinning particle in four-dimensional space-time is a
two-dimensional cylinder that lies in the hyperplane with the space-like
normal. The vector of normal is orthogonal to the direction of
cylinder axis. The direction of cylinder axis is determined by the
momentum. The normal to the hyperplane defines spin. The one-to-one relationship between the particle states and (hyper)cylinders induces the invariant norm on the space of (hyper) cylinders corresponds to the norm. In so doing, the integration over the phase space means summation over the spinning particle world sheets, being cylinders.\footnote{We thank S.L. Lyakhovich for this remark.}

\section{One-particle distribution function}

In this section, we address a question of construction of the one-particle distribution function for a rotating classical non-relativistic ideal gas. The canonical distribution function for a rotating with constant angular velocity macroscopic system proposed in \cite{JWG-1902}. If $\boldsymbol{\Omega}$ denotes the angular velocity and $\theta$ is the temperature, the probability density has the form
\begin{equation}\label{df}
    \rho=\frac{1}{Z}\exp\bigg(-\frac{E-(\boldsymbol{\Omega},\boldsymbol{J})}{\theta}\bigg)\,.
\end{equation}
The quantity $\boldsymbol{J}$ denotes the total (orbital and spinning) angular momentum of all the particles in the system, and the round brackets stand for the scalar product of two vectors. The argument of exponent has sense of energy of the theory in the co-rotating coordinate system \cite{LL}. The quantity $Z$ is the partition function. It is determined from the condition
\begin{equation}\label{pf}
    Z=\int \exp\bigg(-\frac{E-(\boldsymbol{\Omega},\boldsymbol{J})}{\theta}\bigg) d\Gamma_N\,.
\end{equation}
The integration is held over the microscopic states of the system. The subscript $N$ in $d\Gamma_N$ accounts that the invariant measure depends on the number of particles in the system. Throughout the article, we assume that the particles are identical. Then, the volume element $d\Gamma_N$ is the product of $N$ invariant measures on the set of one-particle states (\ref{dg}), with the factor $1/N!$ being included. The ensemble average of the microscopic observable $O$ is determined by the formula
\begin{equation}
    \langle O\rangle =\int O\rho d\Gamma_N\,.
\end{equation}
The averaging principle \cite{JWG-1902} tells us that the quantity $\langle O\rangle$ is the observable value of the quantity $O$.

In the case of ideal gas, the energy and total angular momentum of the system are given by the sum of energies and total angular momenta of individual particles. If the index $a=1,\ldots,N$ labels the particles in the gas, we have the following estimate
\begin{equation}
    E=\sum_{a=1}^{N}\varepsilon_{a}\,,\qquad \boldsymbol{J}=\sum_{a=1}^N\boldsymbol{j}_a.
\end{equation}
The quantities $\varepsilon_a$ and $\boldsymbol{j}{}_a$ are determined in the formula (\ref{J-tot}). The representation of angular momentum in terms of angular variables is given in (\ref{spin-vect}). The absence of interaction between particles implies that the distribution function (\ref{df}) decouples into the product of one-particle distribution functions. The latter reads
\begin{equation}\label{opdf-gen}
    \rho_0=\frac{1}{Z_0}\exp\bigg(-\frac{p^2}{2m\theta}+\frac{(\boldsymbol{\Omega},\boldsymbol{j})}{\theta}\bigg).
\end{equation}
The partition function $Z_0$ is determined by the rule
\begin{equation}\label{opf-gen}
    Z_0=\int \exp\bigg(-\frac{p^2}{2m\theta}+\frac{(\boldsymbol{\Omega},\boldsymbol{j})}{\theta}\bigg)d\Gamma\,,
\end{equation}
with $d\Gamma$ being the phase-space volume element (\ref{dg}). The partition function $Z$ (\ref{pf}) reads
\begin{equation}\label{pf-gen}
    Z=\frac{1}{N!}(Z_0)^N\,.
\end{equation}
The average of the observable $o$ is determined by the formula
\begin{equation}\label{opdf-avg}
    \langle o\rangle =\int o\rho d\Gamma_N\,.
\end{equation}
In the last equality, it is assumed that $o$ is the function of generalized coordinates and generalized momenta of a single particle.

In the current article, we assume that the gas reservoir is the cylinder rotating around its symmetry axis. The radius of the cylinder is denoted by $r$, and the height is $h$. Without loss of generality, we assume that the vector of angular velocity is directed along the third coordinate axis $\boldsymbol{\Omega}=(0,0,\Omega)$. The superscript $3$ in $\Omega^3$ is intentionally omitted because the angular velocity has one non-trivial component. In this setting, the scalar product $(\boldsymbol{\Omega},\boldsymbol{j})$ takes the following form:
\begin{equation}
    (\boldsymbol{\Omega},\boldsymbol{j})=\Omega(x^1p^2-x^2p^1)+\hbar\Sigma\Omega\cos(\beta)\,.
\end{equation}
For one-particle distribution function (\ref{opdf-gen}), we
get
\begin{equation}\label{opdf}
    \rho_0=\frac{1}{Z}\exp\bigg(-\frac{(\tilde{p}^1){}^2+(\tilde{p}^2){}^2+(p^3){}^2}{2m\theta}+\frac{m\Omega^2((x^1){}^2+(x^2){}^2)}{2\theta}+\frac{\hbar\Omega\Sigma\cos(\beta)}{\theta}\bigg)\,.
\end{equation}
The quantities $\tilde{p}^1=p^1-m\Omega x^2$,
$\tilde{p}^2=p^2+m\Omega x^1$ are particle momenta in the co-rotating with the angular velocity $\Omega$ coordinate system. The partition function $Z_0$ reads
\begin{equation}
Z_{0}=\frac{(2\pi m\theta)^{3/2}}{(2\pi\hbar)^3}
\frac{8\pi h\theta^2}{m\hbar\Omega{}^3} \bigg(\exp\bigg(\frac{mr^2\Omega^2}{2\theta}\bigg)-1\bigg)\sinh\bigg(\frac{\hbar\Omega\Sigma}{\theta}\bigg)\,.
\end{equation}
The quantity $Z_0$ is a function of three dimensionless combinations of model parameters,
\begin{equation}\label{xyz}
    x=\frac{\hbar\Omega}{\theta}\,,\qquad y=\frac{m\Omega{}^2r^2}{2\theta}\,,\qquad z=\frac{n}{2S+1}\bigg(\frac
    {2\pi\hbar}{\sqrt{2\pi m\theta}}\bigg)^3.
\end{equation}
($n=N/V$ stands for concentration of the particles). The constant $z$ represents the number of accessible states (per particle). The use of classical statistics suggests that $z\gg1$. Otherwise, the Bose or Fermi distribution functions must be used. The quantities $x$ and $y$ represent the ratios of spin excitation and centrifugal energies to the temperature. In the usual conditions, these parameters are very small (the estimates are given in Section $5$). This makes asymptotic expansions for small $x$ and $y$ useful. We provide these estimates when necessary.

The one-particle distribution
function (\ref{opdf}) is given by the product of three functions
depending on momenta, positions is space, and angular variables.
This means that the dynamics of these degrees of freedom are statistically
independent. For the translational degrees of freedom, we have the
Maxwell-Boltzmann distribution in the field of centrifugal forces,
\begin{equation}\label{opdf-mb}
    \rho_{0,tr}=\frac{1}{Z_{0,tr}}\exp\bigg(-\frac{(\tilde{p}^1){}^2+(\tilde{p}^2){}^2+(p^2){}^2}{2m\theta}+\frac{m\Omega{}^2((x^1){}2+(x^2)^2)}{2\theta}\bigg)\,;
\end{equation}
\begin{equation}\label{pf-mb}
Z_{0,tr}=\frac{(2\pi m\theta)^{3/2}}{(2\pi\hbar)^3}
\frac{4\pi h\theta}{m\Omega{}^2} \bigg(\exp\bigg(\frac{mr^2\Omega^2}{2\theta}\bigg)-1\bigg)\,.
\end{equation}
For the momenta, we have the
Maxwell distribution with temperature $\theta$ in the rotating
coordinate system. The observer that rotates together
with the gas will see the usual Maxwell distribution for the particle
momenta. For the positions, we find the
Boltzmann distribution in the field of centrifugal forces. The distribution function (\ref{opdf-mb}), (\ref{pf-mb}) is well known in the literature \cite{LL}. The function (\ref{opdf}) has an important feature: the distribution law for the translational degrees of freedom have one and the same form 
irrespectively of the value of spin. This means that the usual
Maxwell-Boltzmann distribution does not affect by the presence or
absence of particle spin in classical statistical physics. The conclusion does not apply to the
spinning distribution function, which is spin-sensitive.

As the chiral effects of interest, we should focus on the distribution function for
spinning degree of freedom. Integrating the expression (\ref{opdf}) by
the positions and conjugated momenta, we find
\begin{equation}\label{opdf-sp}
    \rho_{0,sp}(\alpha,\beta)=\frac{x\Sigma\exp(x\Sigma\cos(\beta))}{4\pi\sinh(-x\Sigma)}\,.
\end{equation}
The function $\rho_{0,sp}(\alpha,\beta)$ determines probability
$dw_{0,sp}$ of the spin vector direction (\ref{spin-vect}) into the elementary
solid angle $\cos(\beta)d\alpha d\beta$ by the formula
\begin{equation}
    \phantom{\frac{1}{2}}dw_{0,sp}(\alpha,\beta)=\rho_{0,sp}(\alpha,\beta)\cos(\beta)d\alpha d\beta\,.\phantom{\frac{1}{2}}
\end{equation}
This expression tells us that the distribution for position angle $\alpha$ is uniform. The result is not surprising because the
rotation around the third coordinate axis is a symmetry of the model. The distribution
for the angle $\beta$ is not homogeneous. As is seen from (\ref{opdf-sp}), the
states with $\beta=0$ are more probable than the states with
$\beta=\pi$. This means that the particle spins tend to orient along the
vector of angular momentum. A similar effect is observed under the influence of magnetic field and rotation in accelerators, where the polarization of elementary particles with spin occurs \cite{Hatt}, \cite{Jiang}, \cite{Chen}, \cite{Liu}. The
general conclusion of this paragraph is that the classical theory of spin
describes chiral effects, at least at the level of quality.

Now, it is interesting to describe the polarization of spinning gas
quantitatively. It is convenient to introduce the classical analog
of spin quantum number $s=\cos (\beta)\Sigma$. The quantity
$s$ has the sense of spin projection onto the rotation axis, being
measured in the Planck units. In the classical case, it takes the continuum of values in the
range $s\in[-\Sigma,\leq\Sigma]$. The distribution function
for $s$ reads
\begin{equation}\label{rho-cl}
    \rho_{0,sp}(s)=\frac{2x\exp(x s)}{\sinh(x\Sigma)}\,.
\end{equation}
The function $\rho_{0,sp}$ grows with $s$ monotonously. This means that the particle with the biggest value of spin projection onto the rotation axis are most abundant. In the case of slow rotation, $x\ll 1$, we have the following estimate for the
distribution function (\ref{rho-cl}),
\begin{equation}\label{rho-cl-1}
    \rho_{sp}(s)=\frac{1}{2\Sigma}\bigg(1+x s+\frac{1}{6}x^2(3s^2-\Sigma^2)+o(x^2)\bigg)\,.
\end{equation}
In the first order approximation, the distribution function is linear in $x$. This means that, in the slow rotation limit, the fraction of particles with spin orientation along the angular velocity vector linearly grows with increase of the angular velocity $\Omega$. The formula (\ref{rho-cl-1}) implies that the particles with the orientation of spins along
the direction of angular velocity are $1+2x\Sigma$ times more
probable than the particles with the opposite orientation. This
associates the quantity $\hbar\Omega$ with the energy of spinning degree of freedom excitation.

Another characteristic of distribution (\ref{opdf-sp}) is the
fraction of particles whose spins have an acute angle with the angular velocity vector. The fraction of particles with positive projection of angular momentum is determined by the formula
\begin{equation}\label{w-cl-1}
    w^{+}_{sp}=\frac{2\exp(x\Sigma)-1}{\sinh(x\Sigma)}\,.
\end{equation}
The expression in the right hand side is always positive, so the
majority of vectors of spin of the particle have an acute angle with
the angular velocity. The fraction of particles with positive
projection of spin grows with the increase of rotation angular velocity. The expansion of (\ref{rho-cl-1}) for small values of
$x$ reads
\begin{equation}\label{rho-cl-2}
    w^{+}_{sp}=\frac{1}{2}\bigg(1+x\Sigma-\frac{1}{8}x^3\Sigma^3+o(x^3)\bigg)\,.
\end{equation}
Here, we observe that the fraction of particles with orientation
along the vector of angular momentum grows linearly with the angular
velocity. This confirms the conclusion about the dominating
orientation of spins along the rotation axis even in the classical statistics. From the academic viewpoint, the result means that the classical theory of spin can be
applied for description of chiral effects in the rotating systems,
at least at the level of quality. The question is the magnitude of the chiral effects. We discuss the problem in Section 5.

It is interesting to compute the average value of projection of spin $s$ onto the rotation axis. The quantity $\langle s\rangle$ is determined by the formula (\ref{opdf-avg}). A straightforward computation gives,
\begin{equation}\label{s-avg}
    \phantom{\frac{1}{2}}\langle s\rangle=\Sigma \coth(x \Sigma)-x^{-1}\,.\phantom{\frac{1}{2}}
\end{equation}
The quantity
(\ref{s-avg}) is positive for all the possible values of $x$ (and, hence, $\Omega>0$). The result shows that the projection of spin angular momentum has nonzero expectation value in rotating system. The projection of spin angular momentum of the particle is determined by the formula $\langle j_{sp}\rangle=\hbar\langle s\rangle$. (we omit the subscript $3$ in $j^3_{sp}$ because $\langle\boldsymbol{j_{sp}}\rangle$ has only one independent component). The value of determines the extent of polarization of particle spins by rotation. 
For small values of $x$, we have the following expansion:
\begin{equation}\label{s-avg-1}
    \phantom{\frac{1}{2}}
    \langle s\rangle=\frac{1}{3}x\Sigma-\frac{1}{45}x^3\Sigma^3+o(x^3).
\end{equation}
The leading in $x$ term of this expression determines the polarizability of spinning degree of freedom $\chi$. In the case at
hands, it reads
\begin{equation}\label{chi-avg}
    \phantom{\frac{1}{2}}
    \chi=\frac{\partial\langle j_{sp}\rangle}{\partial \Omega}\Big|_{x=0}=\frac{\hbar\Sigma}{3\theta}.
\end{equation}
Formula (\ref{s-avg}) gives a quasiclassical
estimate of polarizability of rotating spinning gas.

\section{Thermodynamics of spin}

The partition function $Z$ (\ref{pf}) determines
the thermodynamic potential $\Phi$ of the rotating gas by the
standard formula:
\begin{equation}\label{p-z}
    \Phi=-\theta\ln Z.
\end{equation}
The natural variables of the thermodynamic potential
$\Phi$ are the temperature $\theta$, angular
velocity $\Omega$, and particle number $N$. The quantity
(\ref{p-z}) is connected with the internal energy $U$ by the rule
\begin{equation}
    \Phi=U-\langle S\rangle \theta-\Omega \langle J\rangle\,,
\end{equation}
with $\langle J\rangle $ being the average projection of total angular momentum of
all particles in gas onto the rotation axis; $\langle S \rangle$ stands for average entropy. It has the
meaning of the free energy of the system in the co-rotating with
angular velocity $\boldsymbol{\Omega}$ coordinates. The differential
of the quantity $\Phi$ reads
\begin{equation}
    d\Phi=-S d\theta-Jd\Omega+ \mu
    dN\,.
\end{equation}
The last formula suggests that $\Phi$ is the
analog of the Gibbs free energy, where the $PV$-pair of conjugated
variables is replaced by the
$\Omega J$ pair. The thermodynamic
potential $\Phi$ determines the average entropy of the system,
average angular momentum, and average chemical potential
by the rule
\begin{equation}\label{eos}
    \langle S\rangle =-\frac{\partial \Phi}{\partial
    \theta}\,,\qquad \langle J \rangle =-\frac{\partial
    \Phi}{\partial\Omega}\,,\qquad \langle\mu\rangle =\frac{\Phi}{N}\,.
\end{equation}
Once $\Phi$ is determined as the function of natural variables, relations (\ref{eos}) represent equations of state of the rotating gas. 

In terms of the variables $x,y,z$ (\ref{xyz}), the thermodynamic potential of the rotating spinning gas has the following form: 
\begin{equation}\label{phi-z-0}
    \Phi=-\theta N\bigg(\ln\frac{z}{xy}+\ln(\exp(y)-1)+\ln \sinh(x\Sigma)+\ln 2\bigg)\,.    
\end{equation}
In the slow rotation limit $x,y\ll1$, we have estimate,
\begin{equation}\label{phi-z-0-app}
    \Phi=-\theta N\bigg(\ln z+\frac{1}{2}y-\frac{1}{24}y^2+\frac{1}{6}x^2\Sigma^2-\frac{1}{180}x^4\Sigma^4+\ln
2\Sigma+o(y^2)+o(x^4)\bigg)\,.    
\end{equation}
The expressions (\ref{phi-z-0}), (\ref{phi-z-0-app}) are given by the sums of the thermodynamic potential of classical non-rotating gas of scalar particles $-\theta N\ln z$, and the corrections for centrifugal forces and spin. Formulas (\ref{eos}) determine entropy, total angular momentum and chemical potential of the gas. We have the following result for $\langle S\rangle$, $\langle J\rangle$:
\begin{equation}\label{entr-avg}
    \langle S\rangle =N\bigg(\ln\frac{z}{xy} +\ln(\exp(y)-1)-\frac{y\exp(y)}{\exp(y)-1}+\ln \sinh(x\Sigma)-x\Sigma\coth(x\Sigma)+\ln 2-\frac{7}{2}\bigg)\,;  
\end{equation}
\begin{equation}\label{j-avg}
    \langle J\rangle =\frac{N\theta}{\Omega}\bigg(\frac{2y\exp(y)}{\exp(y)-1}+x\Sigma\coth(x\Sigma)-3\bigg)\,.
\end{equation}
We do not give the expression for $\langle\mu\rangle$ as it is determined by (\ref{eos}). In the slow rotation limit $x,y\ll1$, we have the estimates
\begin{equation}\label{entr-avg-0}
    \langle S\rangle =N\bigg(\ln z-\frac{1}{24}y^2-\frac{1}{6}x^2\Sigma^2+\frac{1}{60}x^4\Sigma^2+\ln 2-\frac{3}{2}+o(y^2)+o(x^4)\bigg)\,;  
\end{equation}
\begin{equation}\label{j-avg-0}
    \langle J\rangle =\frac{N\theta}{\Omega}\bigg(y-\frac{1}{6}y^2+\frac{1}{3}x^2\Sigma^2-\frac{1}{45}x^4\Sigma^4+o(y^2)+o(x^4)\bigg)\,.
\end{equation}
The expressions (\ref{entr-avg}), (\ref{j-avg}), (\ref{entr-avg-0}), (\ref{j-avg-0}) represent a mixture of effects of translational and spinning degrees of freedom.

The chiral effects are determined by the spin-dependent part of the thermodynamic potential $\Phi$ (\ref{phi-z-0}), which reads
\begin{equation}
    \phantom{\frac12}\Phi_{sp}=-\theta N \bigg(\ln\coth(x\Sigma)-\ln x+\ln 2\bigg)\,. \phantom{\frac12}
\end{equation}
The entropy, angular momentum, and chemical potential of spinning degree of freedom respectively read
\begin{equation}\label{entr-avg-sp}
    \langle S_{sp}\rangle =N\bigg(\ln \sinh(x\Sigma)-x\Sigma\coth(x\Sigma)-\ln x+\ln 2\bigg)\,;  
\end{equation}
\begin{equation}\label{j-avg-sp}
    \langle J_{sp}\rangle =\frac{N\theta}{\Omega}\bigg(x\Sigma\coth(x\Sigma)-1\bigg)\,,\qquad \langle\mu_{sp}\rangle=-\theta \bigg(\ln\coth(x\Sigma)-\ln x+\ln 2\bigg).
\end{equation}
In the slow rotation limit, we have the estimates
\begin{equation}\label{entr-avg-0-sp}
    \langle S_{sp}\rangle =N\bigg(\ln 2\Sigma-\frac{1}{6}x^2\Sigma^2+\frac{1}{60}x^4\Sigma^2+o(x^4)\bigg)\,;  
\end{equation}
\begin{equation}\label{j-avg-0-sp}
    \langle J_{sp}\rangle =\frac{N\theta}{\Omega}\bigg(\frac{1}{3}x^2\Sigma^2-\frac{1}{45}x^4\Sigma^4+o(x^4)\bigg)\,,\qquad \langle\mu_{sp}\rangle=-\theta \bigg(\ln 2\Sigma+\frac{1}{6}x^2\Sigma^2-\frac{1}{180}x^4+o(x^4)\bigg)\,.
\end{equation}
These results demonstrate that the macroscopic parameters characterising the spinning degree of freedom depend on the angular velocity of rotation. This is another confirmation of presence of chiral effects in classical spinning gas.

The entropy (\ref{entr-avg-0-sp}) of spinning degree of freedom depends on the angular velocity, so the rotation can be converted into the heat and vice versa. The heat capacity of the system at the constant angular velocity reads
\begin{equation}\label{c-sp}
    \langle C_{\Omega,sp}\rangle =N\bigg(1-\frac{x^2\Sigma^2}{\sinh^2(x\Sigma)}\bigg)\,.    
\end{equation}
In the slow rotation limit $x\ll1$, the spinning heat capacity is quadratic in the angular velocity,
\begin{equation}\label{c-x-s}
    \langle C_{\Omega,sp}\rangle =\frac{1}{3}x^2\Sigma^2N\bigg(1-\frac{1}{5}x^2\Sigma^2\bigg)+o(x^4)\,.    
\end{equation}
For the fast rotation, $x\gg1$, we have extreme estimate
\begin{equation}\label{c-x-b}
    \langle C_{\Omega,sp}\rangle=N+o(\frac{1}{x})\,.    
\end{equation}
The formulas have a simple explanation. In the fast rotating system, the spin angular momenta of the particles make small oscillations around the direction of angular velocity vector. The heat capacity of the system is determined by a single oscillatory degree of freedom of each particle. In case of slow rotation, the distribution of spins is almost homogeneous. The excess of particles oriented along the vector of angular velocity is proportional to $x$. The heat capacity is determined by the extra energy required to change the distribution of spins, being proportional to homogeneity of distribution and angular velocity. The combined effect is proportional to $x^2$ in the leading order in $x$. 

\section{Comparison with quantum case}

The quantum mechanics provides the most systematic setting for description of spin. The aim of the section is the comparison of predictions of classical and quantum spin statistics for rotating ideal gas. For the reasons of simplicity, we assume that the gas is non-degenerate, so its temperature is not too low. This allows to restrict ourselves with the Gibbs distribution instead of Bose or Fermi statistics required for the degenerate gases. We show that the classical theory of spin give the correct predictions in the leading order in the Plank constant. The quantum corrections to quasi-classical partition function have the order $x^4$. Given the small value of $x$, we conclude that the classical theory of spin provides the adequate description of chiral effects in non-degenerate gases.

The spin of the quantum particle is the half-integer number $S$. The state of the spinning degree of freedom is determined by the projection of spin angular momentum onto the third coordinate axis, which takes $2S+1$ possible values 
\begin{equation}\label{s-quant}
    -\hbar S\,,-\hbar(S-1)\,,\ldots\,,+\hbar(S-1)\,,+\hbar S.
\end{equation}
The partition function has the following form:
\begin{equation}\label{z-q}
    Z^{\ast}_{sp}=\sum_{s=-S}^{S}\exp(-xs)=\sinh^{-1}\bigg(\frac{1}{2}x\bigg)\sinh\bigg(\frac{2S+1}{2}x\bigg)\,.
\end{equation}
(Asterisk stands for "quantum" expression). 
This leads to the following estimate for the thermodynamic potential of spinning degree of freedom
\begin{equation}\label{}
    \Phi^\ast_{sp}=-\theta N\bigg(\ln\sinh\bigg(\frac{2S+1}{2}x\bigg)-\ln\sinh\bigg(\frac{1}{2}x\bigg)\bigg)\,.
\end{equation}
In the slow rotation limit $x\ll1$, the expansion for the quantity $\Phi_{sp}^\ast$ reads
\begin{equation}\label{}
    \Phi^\ast_{sp}=-\theta\bigg(\ln(2S+1)+\frac{1}{6}x^2\Sigma^2-\frac{1}{180}x^4\Sigma^2(\Sigma^2+\frac{1}{2})+o(x^4)\bigg)\,.
\end{equation}
The comparison of this formula with (\ref{phi-z-0-app}) shows two differences. The leading term in brackets is $\ln (2S+1)$, not $\ln 2\Sigma$. The shift caused by the selected normalization of the invariant of the measure on the phase space. The shift corresponds to redefinition of entropy of initial state. It does not affect the thermodynamics of the model. The first non-trivial contribution $1/6x^2\Sigma^2$ is one and the same in both the expansions. The quantum corrections connected with the discrete change of spin states have the order $O(x^4)$. Given the definition of the dimensionless parameter $x$, this means that the spin corrections in thermodynamics potential are proportional 
to the second order of Planck constant 
$\hbar^2$. The classical theory of spin and classical statistical mechanics gives the correct estimate for the spin effects in the leading order in $\hbar$. The quantum corrections to the thermodynamic potential are proportional to the fourth order of the Planck constant $\hbar^4$. 

Our final remark concerns the precision of predictions of classical theory of spin. The magnitude of chiral effects is determined by the value of dimensionless parameter $x$ (\ref{xyz}). In the most favorable case, the temperature $\theta$ has the lowest possible value, and the angular velocity $\Omega$ has the maximal possible value. The lowest temperature of the classical gas is the temperature of degeneration $\theta^\ast$\footnote{For the Bose gas. For the Fermi gas, the expression should be multiplied by $3^{\frac23}\pi^{\frac53}[\zeta(3/2)]^{\frac23}\approx 60$.}
\begin{equation}
    \theta^\ast=\frac{\hbar^2}{2\pi m}\bigg(\frac{n}{(2S+1)\zeta(3/2)}\bigg)^{\frac23}\,,\qquad \zeta(3/2)=2.612\cdots.
\end{equation}
The maximal possible angular velocity is limited by competition between centrifugal and chiral effects. Both the effects have equal magnitude if $x=y$ (\ref{xyz}). The critical angular velocity $\Omega^\ast$ has the following form: 
\begin{equation}
    \Omega^\ast=\frac{2\hbar}{mr^2}.
\end{equation}
In terms of the variables $\theta^\ast$, $\Omega^\ast$, the quantity $x$ admits the following representation
\begin{equation}
    x=\frac{\Omega}{2\pi\Omega^\ast}\bigg(\frac{(2S+1)\zeta(3/2)}{n}\bigg)^{2/3}
\end{equation}
Here, $\Omega$ is the angular velocity, and $n$ is the concentration. Assuming that the cylinder height is equal to the radius, $r=h$, and $\Omega=\Omega^\ast$, we obtain
\begin{equation}
    x^\ast=\bigg(\frac{(2S+1)\zeta(3/2)}{4\pi^2N}\bigg)^{2/3}\,.
\end{equation}
For $N=10^{17}$ (the number of gas molecules in a cubic millimeter under the normal conditions), and $S=0$, we get $x^\ast\approx 10^{-12}$. This is a very small quantity. The dimensionless intensive quantities (specific entropy, specific heat capacity, specific angular momentum, and chemical potential) have the order $x$ or $x^2$, i.e. $10^{-12}$ or $10^{-24}$. The magnitude of quantum corrections macroscopic parameters have the order $x^3\approx 10^{-36}$ or $x^4\approx 10^{-48}$. The relative precision of quasi-classical estimates is of order $10^{-24}$. This corresponds to a very high precision of quasi-classical estimates. This means that the classical statistical mechanics and classical theory provide effective setting for description of chiral effects in non-degenerate gas of spinning particles.

\section{Results}

In this article, we have considered a statistical mechanics and
thermodynamics of classical non-relativistic gas of spinning
particles. In the applied quasi-classical approach, the microscopic gas
particles are considered as point objects in Euclidean space that
are equipped with internal angular momentum. The latter is treated as
spin. The vector of spin is normalized. The norm is determined by
the spin of particle. In contrast to the quantum case, the vector
classical spin runs over a continuous set of states. The phase space
of the particle is shown to be $\mathbb{R}^6\times\mathbb{S}^2$. The
invariant measure on the phase space is given by the product of
invariant measures on the space of translational and spinning
degrees of freedom.

Using the canonical formalism of statistical mechanics of rotating
systems, we have constructed the one-particle distribution function.
It is given by the product of translational and spinning
contributions. The translational degrees of freedom subjects to the
usual Maxwell-Boltzmann distribution, while the spin degree of
freedom has its own distribution law. As we have observed, there is
a uniform distribution of spin directions in the orthogonal to
rotation axis plane. For the projections onto the rotation axis, we
observed the excess of particles with spins oriented against the angular
velocity vector. This fact demonstrates that the majority of the
particles have the orientation in the direction of angular velocity. This excess is quantitatively described.

We have construed partition function and thermodynamic potential of
rotating gas. We have found the equations of state, entropy, and heat
capacity at constant angular velocity. The macroscopic quantities are characterized by two dimensionless parameters: the ratio of
specific centrifugal energy (per particle) to the energetic temperature,
and the ratio of spin excitation energy to energetic
temperature. The spin effects dominate if the second parameter is
the biggest. A simple estimate shows that spin effects dominate at
small angular velocities irrespectively of temperature. The critical angular velocity is the biggest for light particles. However, the
size of chiral effects is small in all the cases. The typical value of chiral corrections have the order $10^{-12}$ in the optimal conditions.

The results of the article show that the predictions of classical
and quantum theory for chiral effects coincide in the first non-trivial
order in angular velocity. The quantum contributions are proportional to the fourth order of the ratio of spin excitation energy to the energetic
temperature. This quantity is small, so the classical theory becomes a good approximation for 
rotating gases of spinning particles above the degeneracy temperatures. The classical estimates have one important advantage: the spin changes continuously, so they valid for the particles with all the possible values of spin. This makes the classical theory of
spinning particles provides a universal formalism for description of chiral effects in rotating
systems. 

There are two directions of further research. First, the statistical mechanics of rotating systems can be applied in the class of relativistic particles. The result is the generalization of the well-known Maxwell-Juttner
distribution for classical particles with spin. Second, the models with the external electromagnetic or gravitational fields are of interest. Here, the (possible) phase transitions under the combined action of the external field and rotation represent the most intriguing aspect of the issue. We are planning to address these problems in further studies.

\section*{Acknowledgments}
The authors thank A.A. Sharapov, P.O. Kazinski, and Yu.V. Brezhnev for valuable discussions of this work. The work was supported by the RSF project 21-71-10066.

\renewcommand\refname{Bibliography}

\end{document}